\documentclass[twocolumn,spanish,english]{revtex4-1}
\usepackage[T1]{fontenc}
\usepackage[latin9]{inputenc}
\usepackage{amsmath}
\usepackage{amssymb}
\usepackage{graphicx}
\usepackage{esint}
\usepackage{babel}
\addto\shorthandsspanish{\spanishdeactivate{~<>}}

\begin{document}
\selectlanguage{spanish}%
\global\long\def\abs#1{\left|#1\right|}
\global\long\def\ket#1{\left|#1\right\rangle }
\global\long\def\bra#1{\left\langle #1\right|}
\global\long\def\half{\frac{1}{2}}
\global\long\def\partder#1#2{\frac{\partial#1}{\partial#2}}
\global\long\def\comm#1#2{\left[#1,#2\right]}
\global\long\def\vp{\vec{p}}
\global\long\def\vpp{\vec{p}'}
\global\long\def\dt#1{\delta^{(3)}(#1)}
\global\long\def\Tr#1{\textrm{Tr}\left\{  #1\right\}  }
\global\long\def\Real#1{\mathrm{Re}\left\{  #1 \right\}  }
\global\long\def\Braket#1{\langle#1\rangle}
\global\long\def\escp#1#2{\left\langle #1|#2\right\rangle }
\global\long\def\elmma#1#2#3{\langle#1\mid#2\mid#3\rangle}
\global\long\def\ketbra#1#2{|#1\rangle\langle#2|}

\selectlanguage{english}%

\title{A study of decoherence effects in the Stern-Gerlach experiment using
matrix Wigner functions}

\author{Pablo Gomis}

\author{A. Pérez}

\affiliation{Departamento de F\'{í}sica Teórica and IFIC, Universidad de Valencia-CSIC,
Dr. Moliner 50, 46100-Burjassot, Spain}
\begin{abstract}
We analyze the Stern-Gerlach experiment in phase space with the help
of the matrix Wigner function, which includes the spin degree of freedom.
Such analysis allows for an intuitive visualization of the quantum
dynamics of the apparatus. We include the interaction with the environment,
as described by the Caldeira-Leggett model. The diagonal terms of
the matrix provide us with information about the two components of
the state, that arise from interaction with the magnetic field gradient.
In particular, from the marginals of these components, we obtain an
analytical formula for the position and momentum probability distributions
in presence of decoherence, that show a diffusive behavior for large
values of the decoherence parameter. These features limit the dynamics
of the present model. We also observe the decay of the non-diagonal
terms with time, and use this fact to quantify the amount of decoherence,
from the norm of those terms in phase space. From here, we can define
a decoherence time scale, which differs from previous results that
make use of the same model.
\end{abstract}
\maketitle

\section{Introduction}

The Stern-Gerlach (SG) experiment is a cornerstone in quantum mechanics.
It showed, for the first time, direct evidence for the discretization
of the spin states of the electron, by analyzing the motion of Silver
atoms through a magnetic field gradient \cite{Gerlach1922}. Most
textbooks make continuous use of the SG, as a simple way to illustrate
the quantum measurement process, since the electron spin only involves
a two-dimensional Hilbert space. 

A consistent description of the SG experiment needs, obviously, to
be quantum, even though one can make an introduction based on a semiclassical
description, using a spin-dependent force that gives rise to a ``trajectory''
that depends on the initial spin state. The full quantum treatment
reveals a richer dynamics, as it leads to entanglement between the
spin and spatial degrees of freedom \cite{1999PhLA..259..427R,2003EL.....61..148C,2005EPJD...36..235T,2010APS..MARX42010V,0143-0807-26-4-012}. 

In this paper, we perform a phase space analysis of the SG device,
including the interaction with the environment. This interaction will
be described by the Caldeira-Leggett model \cite{PhysRevLett.46.211}.
In this respect, the starting point is similar to the analysis in
\cite{1995PhyA..220..563V}. However, our description is based on
the use of the Wigner function (WF) \cite{PhysRev.40.749}. Wigner
functions have proven to be a powerful tool in physics, and can be
used as an alternative formulation of quantum phenomena, including
their dynamics. The particular features of the phase space description
make it particularly advantageous in some situations, for instance
recognizing the quantum features of states, or dealing with decoherence
scenarios. In the WF, interference effects manifest in a clear way
\cite{Lee2011,PhysRevA.74.042323,Hillery1984121}. 

In order to include the spin degree of freedom, one needs to extend
the usual definition of the WF. A common prescription in the literature
is the use of a matrix valued WF \cite{PhysRevA.56.1205}, where the
spin indices give rise to the matrix elements. Such description has
some advantages when dealing with a particle subject to a spin-dependent
force, since some effects like the spin precession, or motion that
depends on the spin component, are better visualized with respect
to a fixed spin basis. Examples of this description are a previous
analysis of the Stern-Gerlach experiment without including decoherence
effects \cite{Utz2015SG}, the study of entangled vibronic quantum
states of a trapped atom \cite{PhysRevA.56.1205}, or the reconstruction
of the fully entangled quantum state for the cyclotron and spin degrees
of freedom of an electron in a Penning trap \cite{Massini2000}.

As we will show, the phase space description provides a clear visualization
of the SG phenomenology. First, the diagonal terms show the motion
of the two components of the quantum state (corresponding to spin
up or down along the gradient direction). By considering an initial
Gaussian state with arbitrary spin direction, we can obtain the marginals
from the Wigner function, which describe the appropriate probability
distribution function (PDF) of position or momentum \cite{Schleich2001}.
Each of these PDF have a Gaussian shape with a center and width which
are modified by the interaction with the environment, and give valuable
information about the state evolution. On the other hand, the out
of diagonal terms can be used to describe the effect of decoherence,
which manifest into a damping of the norm associated to these terms.
We use this norm as a figure of merit to quantify the amount of decoherence
experienced by the system, as a function of the parameter $\gamma$
that quantifies the strength of the coupling with the environment.
As a result, we obtain a decoherence time which scales as $\gamma^{-1/5}$,
in contrast with previous results that claimed a $\gamma^{1/3}$ scaling
\cite{1995PhyA..220..563V}. We argue that our result is more realistic,
as it implies that a larger decoherence parameter manifests into a
shorter decoherence time scale.

The rest of this paper is organized as follows. In Sect. II, we solve
the equations that govern the evolution of the matrix WF for the SG
device, when the master equation based on the Caldeira-Leggett model
is introduced to describe the environment. By evaluating the marginals
of the diagonal elements, we obtain the position and momentum PDFs,
and we analyze some limiting situations. Sect. III is devoted to the
analysis of our results in a standard setup of the SG experiment.
In particular, we study the damping of the off-diagonal terms, and
make use of this to define a decoherence time scale. Finally, we discuss
the validity of the model to describe decoherence phenomena for the
above setup. Our conclusions are presented in Sect. IV, while some
cumbersome expressions have been relegated to the Appendix.

\section{\label{edo}Dynamics of matrix Wigner functions including interaction
with the environment}

The behavior of a spin 1/2 neutral particle under the action of a
magnetic field gradient, including the influence of decoherence effects,
will be studied in this section. Particles entering the SG apparatus
will move initially along the tube, defined as the $x$ axis. The
geometry of the magnetic field can be described by a dependence of
the form 
\begin{equation}
\vec{B}(x,y,z)=\eta y\vec{j}+(B_{0}-\eta z)\vec{k,}
\end{equation}
which contains a uniform part $B_{0}$, and a gradient of magnitude
$\eta$ on the plane orthogonal to the $x$ axis. Notice that both
position dependent terms in the latter equation are necessary in order
to satisfy $\vec{\nabla}\cdot\vec{B}=0$ and $\vec{\nabla}\times\vec{B}=0$.
However, it can be shown that the effect of the magnetic field contribution
along the $y$ direction causes fast oscillations due to Larmor precession,
which can be averaged out. Following \cite{Platt1992}, we neglect
this contribution (see also \cite{0143-0807-26-4-012}). In this way,
in absence of decoherence effects, the problem can be effectively
factorized as the free propagation along the $x$ and $y$ axis, and
the nontrivial motion corresponding to the $z$ coordinate, which
can be described by the Hamiltonian 
\begin{equation}
H=\frac{p^{2}}{2m}+\frac{g_{s}\mu_{B}}{2}(B_{0}-\eta z)\sigma_{z}\equiv\frac{p^{2}}{2m}+\lambda(B_{0}-\eta z)\sigma_{z},\label{hamiltonian}
\end{equation}
where $p$ is the canonical conjugate momentum for $z$, and $m$
and $g_{s}$ are the mass and gyromagnetic ratio of the particle,
respectively. In Eq. (\ref{hamiltonian}), $\sigma_{z}$ is the third
Pauli matrix, and $\mu_{B}$ the Bohr's magneton.

In order to account for decoherence effects, we assume that they are
described by the Caldeira-Leggett master equation \cite{Breuer2007},
which accounts for those effects in the system via collisions with
a thermal bath of particles. In the position and spin representation,
with the Hamiltonian (\ref{hamiltonian}), the master equation can
be written as \cite{1995PhyA..220..563V} 
\begin{eqnarray}
\frac{\partial\rho_{\alpha\beta}(z,z',t)}{\partial t}= &  & \left[\frac{i\hbar}{2m}\left(\frac{\partial^{2}}{\partial z^{2}}-\frac{\partial^{2}}{\partial z'^{2}}\right)+\frac{i\lambda B_{0}}{\hbar}(\alpha-\beta)\right.\nonumber \\
 &  & -\frac{i\eta\lambda}{\hbar}(\alpha z-\beta z')-\gamma(z-z')\left(\frac{\partial}{\partial z}-\frac{\partial}{\partial z'}\right)\nonumber \\
 &  & \left.-\frac{D}{\hbar^{2}}(z-z')^{2}\right]\rho_{\alpha\beta}(z,z',t).\label{edo-1}
\end{eqnarray}
In the latter equation, $\rho_{\alpha\beta}(z,z',t)\equiv\bra{z,\alpha}\rho(t)\ket{z',\beta}$
are the matrix elements of the density operator $\rho(t)$ representing
the particle state, at a given time $t$, on the basis $\{\ket{z,\alpha}\equiv\ket z\otimes\ket{\alpha}\}$
, where $\ket z/z\in\mathbb{R}$ is the eigenbasis of the position
operator, and $\{\ket{\alpha}\}$ is a fixed basis in spin space.
We find it convenient to choose the eigenstates of $S_{z}$ ($\ket{S_{z}=+\hbar/2}=\ket +$,
$\ket{S_{z}=-\hbar/2}=\ket -$). Finally, $\gamma$ is the damping
rate of the system in the environment. The coefficient $D$ is defined
as $D=2m\gamma k_{B}T$, with $k_{B}$ the Boltzmann's constant, and
$T$ the temperature of the environment.

As discussed in the Introduction, the analysis of the dynamics of
this model will be presented on phase space, with the help of Wigner
matrices 
\begin{equation}
W_{\alpha\beta}(z,p,t)=\frac{1}{2\pi\hbar}\int_{-\infty}^{\infty}ds\ e^{-i\frac{p\cdot s}{\hbar}}\Braket{z+\frac{s}{2},\alpha|\rho(t)|z-\frac{s}{2},\beta},\label{wignerdef}
\end{equation}
where $W_{\alpha\beta}(z,p)$ are the spin matrix elements of the
Wigner function in the above-mentioned $S_{z}$ base. The matrix WF
has, among others, the following properties: 
\begin{enumerate}
\item One has 
\begin{equation}
W_{\beta\alpha}(z,p,t)=W_{\alpha\beta}^{*}(z,p,t),
\end{equation}
which implies that the matrix WF is Hermitian. 
\item The normalization condition becomes
\begin{equation}
\sum_{\alpha}\int_{-\infty}^{\infty}\int_{-\infty}^{\infty}W_{\alpha\alpha}(z,p,t)dzdp=1.
\end{equation}

\item The marginal distributions of (\ref{wignerdef}) are related to matrix
elements of the density operator. In particular, for the diagonal
components we have 
\begin{equation}
\int_{-\infty}^{\infty}W_{\alpha\alpha}(z,p,t)dp=\langle z,\alpha|\rho(t)|z,\alpha\rangle\equiv f^{(\pm)}(z,t),\label{margz}
\end{equation}
with $\alpha=\pm$, where $f^{(\pm)}(z,t)$ represents the position
PDF for the particle. In a similar way, the marginal over the position
variable 
\begin{equation}
\int_{-\infty}^{\infty}W_{\alpha\alpha}(z,p,t)dz=\langle p,\alpha|\rho(t)|p,\alpha\rangle\equiv g^{(\pm)}(p,t)\label{margp}
\end{equation}
represents the momentum PDF, with $\ket{p,\alpha}\equiv\ket p\otimes\ket{\alpha}$,
$\{\ket p/p\in\mathbb{R}\}$ being the eigenstates of the momentum
operator.
\end{enumerate}
\noindent Using the definition Eq. (\ref{wignerdef}) and the dynamics
of the density matrix Eq. (\ref{edo-1}), one can obtain the corresponding
differential equations for the Wigner function, given as follows 
\begin{eqnarray}
\frac{\partial W_{d}^{(\pm)}(z,p,t)}{\partial t}= &  & -\frac{p}{m}\frac{\partial W_{d}^{(\pm)}(z,p,t)}{\partial z}+D\frac{\partial^{2}W_{d}^{(\pm)}(z,p,t)}{\partial p^{2}}\nonumber \\
 &  & +\gamma\frac{\partial(pW_{d}^{(\pm)}(z,p,t))}{\partial p}\nonumber \\
 &  & \mp\eta\lambda\frac{\partial W_{d}^{(\pm)}(z,p,t)}{\partial p},\label{diagedo}
\end{eqnarray}
\begin{eqnarray}
\frac{\partial W_{od}(z,p,t)}{\partial t}= &  & -\frac{p}{m}\frac{\partial W_{od}(z,p,t)}{\partial z}+D\frac{\partial^{2}W_{od}(z,p,t)}{\partial p^{2}}\nonumber \\
 &  & +\gamma\frac{\partial(pW_{od}(z,p,t))}{\partial p}\nonumber \\
 &  & \pm\frac{2i\lambda(B_{0}+\eta z)W_{od}(z,p,t)}{\hbar},\label{offedo}
\end{eqnarray}
where $W_{d}^{(\pm)}(z,p,t)$ stands for the diagonal elements of
the Wigner function, with the upper sign corresponding to $W_{++}(z,p,t)$,
and the lower sign to $W_{--}(z,p,t)$. We also defined $W_{od}(z,p,t)\equiv W_{+-}(z,p,t)$,
which implies that $W_{-+}(z,p,t)=W_{od}^{*}(z,p,t)$, according to
property 1 above.

\subsection{\label{solutions} General solution of the differential equations}

To solve the system of equations (\ref{diagedo},\ref{offedo}) a
Fourier transform is performed over both $z$ and $p$ \cite{1995PhyA..220..563V}.
Once the equations are solved, the Wigner functions are retrieved
by performing the inverse Fourier transform. 

Assuming that the incident particle is described by a polarized Gaussian
beam with spin $\ket n=a\ket ++b\ket -$ (with $|a|^{2}+|b|^{2}=1$),
the initial density operator $\rho(0)$ can be written as
\begin{equation}
\rho(0)=\ket{\psi}\bra{\psi}\otimes\ket n\bra n,\label{initialrho}
\end{equation}
with the wave function that represents the initial state $\ket{\psi}$
in position space defined as 
\begin{equation}
\psi(z)=\frac{1}{(\pi\sigma^{2})^{1/4}}e^{-\frac{z^{2}}{2\sigma^{2}}}.
\end{equation}

From the initial state Eq. (\ref{initialrho}) one can obtain the
matrix Wigner function at $t=0$, which can be written as
\begin{equation}
W(z,p,0)=\begin{pmatrix}|a|^{2} & ab^{*}\\
a^{*}b & |b|^{2}
\end{pmatrix}W_{i}(z,p),
\end{equation}
where 
\begin{equation}
W_{i}(z,p)=\frac{e^{-\frac{\sigma^{2}p^{2}}{\hbar^{2}}-\frac{z^{2}}{\sigma^{2}}}}{\pi\hbar}\label{initial}
\end{equation}
represents the Wigner function for a spinless Gaussian state, and
$\sigma$ is the Gaussian width of the particle's spatial PDF.

Solving the differential equation (\ref{diagedo}) by the method commented
above, with the help of the initial condition (\ref{initial}), one
finds the general solution for the diagonal elements. After some algebra,
they can be written as follows
\begin{equation}
W_{d}^{(\pm)}(z,p,t)=\frac{\gamma^{2}m\sigma}{\pi\sqrt{G(\tau)}}e^{-\frac{F(z,p,\tau)}{G(\tau)}},\label{wdiag}
\end{equation}
and we have defined the new variable $\tau\equiv\gamma t$. The functions
$F(z,p,\tau)$ and $G(\tau)$ are defined in the Appendix. In these
functions we introduced the notations
\begin{equation}
z_{c}=\frac{\eta\lambda\left(\tau+e^{-\tau}-1\right)}{\gamma^{2}m},\label{zc}
\end{equation}
\smallskip{}
\begin{equation}
p_{c}=\frac{\eta\lambda\left(1-e^{-\tau}\right)}{\gamma}.\label{pc}
\end{equation}
The role played by $z_{c}$ and $p_{c}$ will be discussed in the
next Section.

Using the same procedure for the differential equation (\ref{offedo}),
the solution for the off-diagonal elements is also found. The resulting
expression is lengthy so that, in order to express it in a more compact
way, we introduced the functions $C_{i}(\tau)$ ($i=1,2,3,4,5,6$),
that can be found in the Appendix. The off-diagonal elements can finally
be written as
\begin{widetext}
\begin{eqnarray}
W_{od}(z,p,t)=\frac{e^{\left[\frac{2iB_{0}\lambda t}{\hbar}+C_{1}-\frac{(-2C_{2}C_{3}\hbar+C_{4}C_{5}\hbar-iC_{5}z\hbar+2iC_{3}p)^{2}}{4C_{3}\hbar^{2}\left(4C_{6}C_{3}-C_{5}^{2}\right)}+\frac{(z+iC_{4})^{2}}{4C_{3}}\right]}}{2\pi\hbar\sqrt{\left(4C_{3}C_{6}-C_{5}^{2}\right)}} &  & ,\label{woff}
\end{eqnarray}
 where we have omitted, for simplicity, the dependence of $C_{i}(\tau)$
on the variable $\tau$. The matrix Wigner function takes the following
form 
\begin{equation}
W(z,p,t)=\begin{pmatrix}|a|^{2}\ W_{d}^{(+)}(z,p,t) & ab^{*}\ W_{od}(z,p,t)\\
a^{*}b\ W_{od}^{*}(z,p,t) & |b|^{2}\ W_{d}^{(-)}(z,p,t)
\end{pmatrix}.
\end{equation}
As we discuss below, the diagonal terms in $W(z,p,t)$ describe the
behavior of the particles in phase space, and the off-diagonal terms
represent the coherence of the state. 
\end{widetext}

\subsection{\label{marginals} Marginals of the Wigner function: position and
momentum PDFs}

The Wigner function (for a spinless particle) can not be associated
with a probability distribution in phase space: In fact, it is referred
to as a \textit{quasi-probability distribution}, and may even take
negative values. This had to be expected from first principles, given
the incompatibility of the position and momentum observables in quantum
mechanics. One can, however, obtain the PDF corresponding to the particle
position by integrating over the momentum variable, and vice-versa,
as described in the previous Sect. Eq. (\ref{margz}) can be integrated,
with the result
\begin{equation}
f^{(\pm)}(z,t)=\frac{e^{-\frac{(z\mp z_{c})^{2}}{\sigma_{z}^{2}}}}{\sqrt{\pi}\sigma_{z}},\label{probz}
\end{equation}
where 
\begin{equation}
\sigma_{z}^{2}=\frac{2D\left(2\tau+4e^{-\tau}-e^{-2\tau}-3\right)}{\gamma^{3}m^{2}}+\frac{\hbar^{2}\left(1-e^{-\tau}\right)^{2}}{\gamma^{2}m^{2}\sigma^{2}}+\sigma^{2}
\end{equation}
 is the squared width of the position distribution. Integration in
Eq. (\ref{margp}) leads to the momentum PDF:
\begin{equation}
g^{(\pm)}(p,t)=\frac{e^{-\frac{(p\mp p_{c})^{2}}{\sigma_{p}^{2}}}}{\sqrt{\pi}\sigma_{p}},\label{probp}
\end{equation}
with 
\begin{equation}
\sigma_{p}^{2}=\frac{2D\left(1-e^{-2\tau}\right)}{\gamma}+\frac{\hbar^{2}e^{-2\tau}}{\sigma^{2}}
\end{equation}
 giving the squared width of the momentum distribution. 

The above results for the marginals, Eqs. (\ref{probz},\ref{probp})
clearly show that the diagonal components of the Wigner matrix correspond
to Gaussian distributions in phase space $(z,p)$ which center $(\pm z_{c},\pm p_{c})$
and width depend both on time, and on the rest of parameters of the
problem, including the decoherence constants $\gamma$ and $D$. 

Let us notice the following properties of these quantities:
\begin{enumerate}
\item By differentiating Eqs. (\ref{zc}) and (\ref{pc}) one readily obtains
\begin{equation}
\frac{dz_{c}}{dt}=\frac{p_{c}}{m},
\end{equation}
\begin{equation}
\frac{dp_{c}}{dt}=\eta\lambda-\gamma p_{c}.
\end{equation}
 which can be easily identified as the classical equations of motion
for a particle subject to a constant force, plus a friction term.
These equations allow us to describe the motion of the center of the
two Gaussians using a semiclassical framework (especially if we neglect
the interaction with the environment, as done in most textbooks).
\item Let us consider the limit $\gamma t\ll1$. Performing a Taylor expansion
gives 
\begin{equation}
z_{c}\simeq\frac{\eta\lambda t^{2}}{2m},\,\,\,\sigma_{z}\simeq\sigma+\frac{\hbar^{2}}{2m^{2}\sigma^{3}}t^{2},
\end{equation}
\begin{equation}
p_{c}\simeq\eta\lambda t,\,\,\,\sigma_{p}\simeq\frac{\hbar}{\sigma}+\frac{\hbar}{\sigma}(\frac{2D\sigma^{2}}{\gamma\hbar^{2}}-1)\gamma t.\label{pcsigmapsmallt}
\end{equation}
If we further neglect the last term in Eq. (\ref{pcsigmapsmallt}),
the above results can be easily interpreted as the action of a constant
force on the particle, and agree with the ones expected for the experiment
in a decoherence free environment \cite{Platt1992}.
\item In the opposite limit (i.e., when $\gamma t\gg1$) we can approximate
\begin{equation}
z_{c}\simeq\frac{\eta\lambda}{\gamma m}t,\,\,\,\sigma_{z}\simeq\sqrt{\frac{8k_{B}Tt}{\gamma m}},
\end{equation}
\begin{equation}
p_{c}\simeq\frac{\eta\lambda}{\gamma},\,\,\,\sigma_{p}\simeq\sigma_{p}^{\infty}\equiv\frac{2D}{\gamma}=\sqrt{4mk_{B}T}\label{pcsigmapinf}
\end{equation}
which leads to a limiting value of the momentum center (and width
as well). This is a well known effect in classical mechanics, that
appears under the action of a friction force, and will play an important
role in our analysis when large values of $\gamma$ are involved.
We also obtain $\sigma_{z}\propto\sqrt{t}$, i.e. the characteristic
behavior of a diffusive regime. Fig. \ref{figzcandpc} features the
evolution in time of the center of the PDF (both in position and momentum),
using the values of the parameters as defined in the next Sect. For
low values of $\gamma$, the center $z_{c}$ of the position PDF first
grows quadratically, and then it does linearly. The growth time scale
is dictated by $\gamma^{-1}$, so that for very large values of this
parameter $z_{c}$ remains close to zero. On the other hand, the center
$p_{c}$ of the momentum PDF grows linearly for moderate values of
$\gamma$, but approaches a constant value as $\gamma$ is increased,
the asymptotic limit being inversely proportional to the decoherence
parameter. In Fig. \ref{figsigmazandsigmap} we have plotted the width
of the position and momentum distributions as a function of time.
The position width $\sigma_{z}$ grows quadratically for small values
of $t$ (as compared to $\gamma^{-1}$), whereas the regime $\sigma_{z}\propto\sqrt{t}$
corresponds to late times. The transition can only be seen for $\gamma=1$
on this Figure, given the $\gamma^{-1}$ scaling. As for the plots
representing $\sigma_{p}$, one can check that the ratio $\frac{2D\sigma^{2}}{\gamma\hbar^{2}}$
is of the order $\thicksim10^{13}$ (notice that this ratio is independent
of $\gamma$). Eq. (\ref{pcsigmapsmallt}) then predicts a fast increase
of $\sigma_{p}$ even at early times, as it is clearly observed from
these plots. This magnitude will then reach the asymptotic value $\sigma_{p}^{\infty}$
also on the same $\gamma^{-1}$ time scale, which can only be appreciated,
in that Figure, for the $\gamma=10^{10}$ case. 
\begin{figure}
\includegraphics[width=8cm]{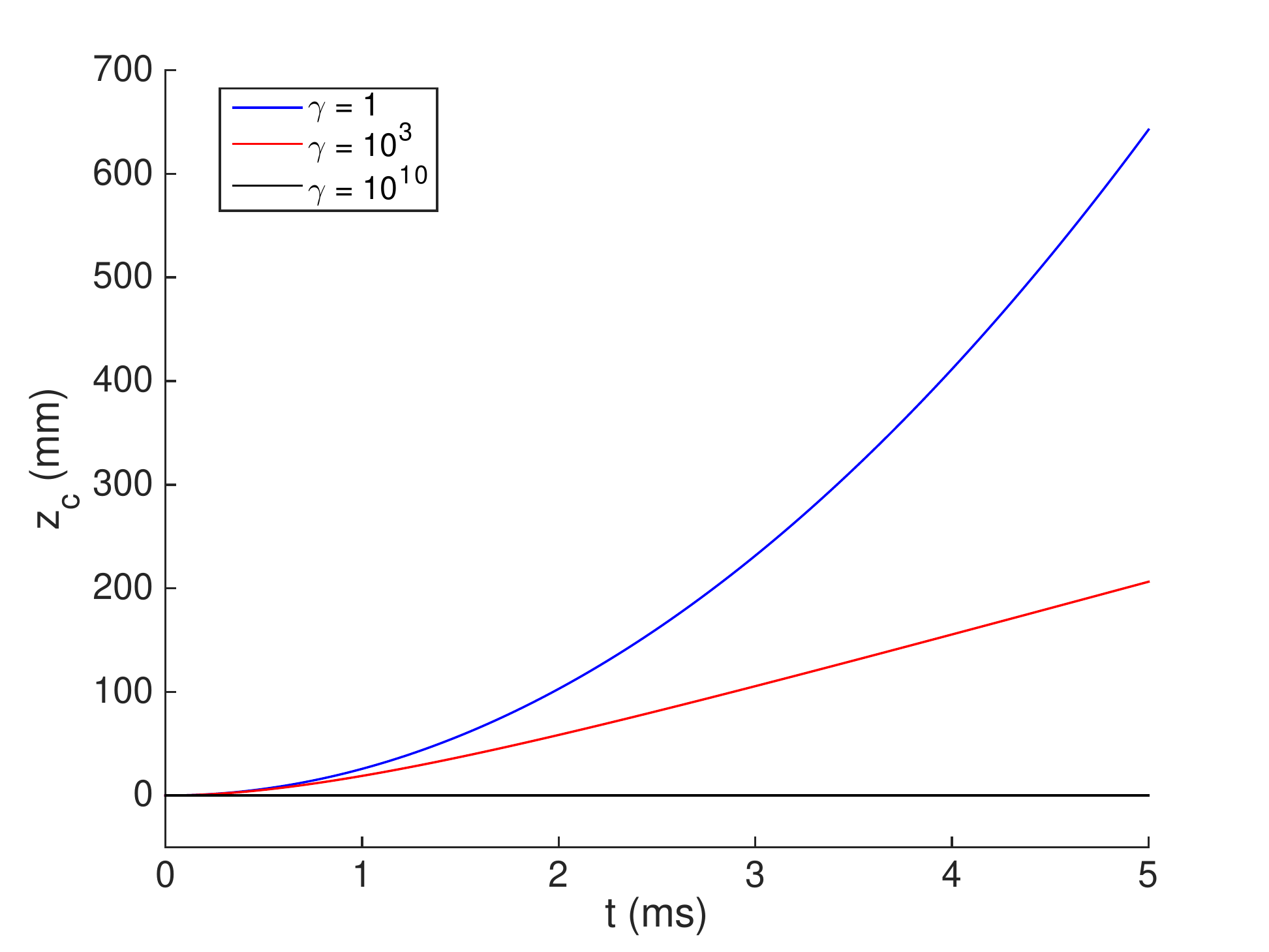}

\includegraphics[width=8cm]{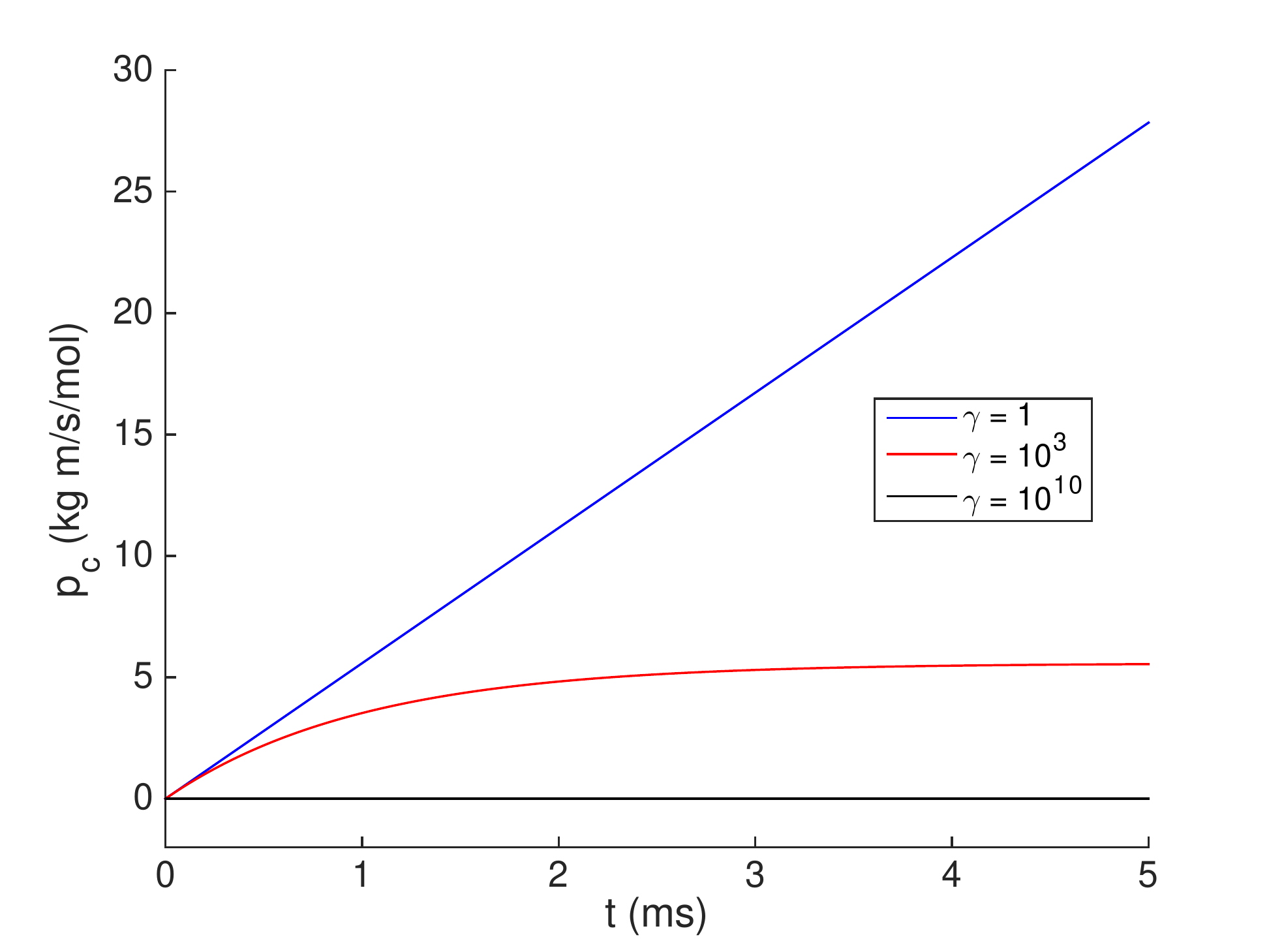}

\protect\caption{(Color online) Plots of the center of the position (up) and momentum
(down) PDFs, as given by Eqs. (\ref{zc}) (\ref{pc}), respectively,
for three values of the parameter $\gamma$. }

\label{figzcandpc}
\end{figure}
\begin{figure}
\includegraphics[width=8cm]{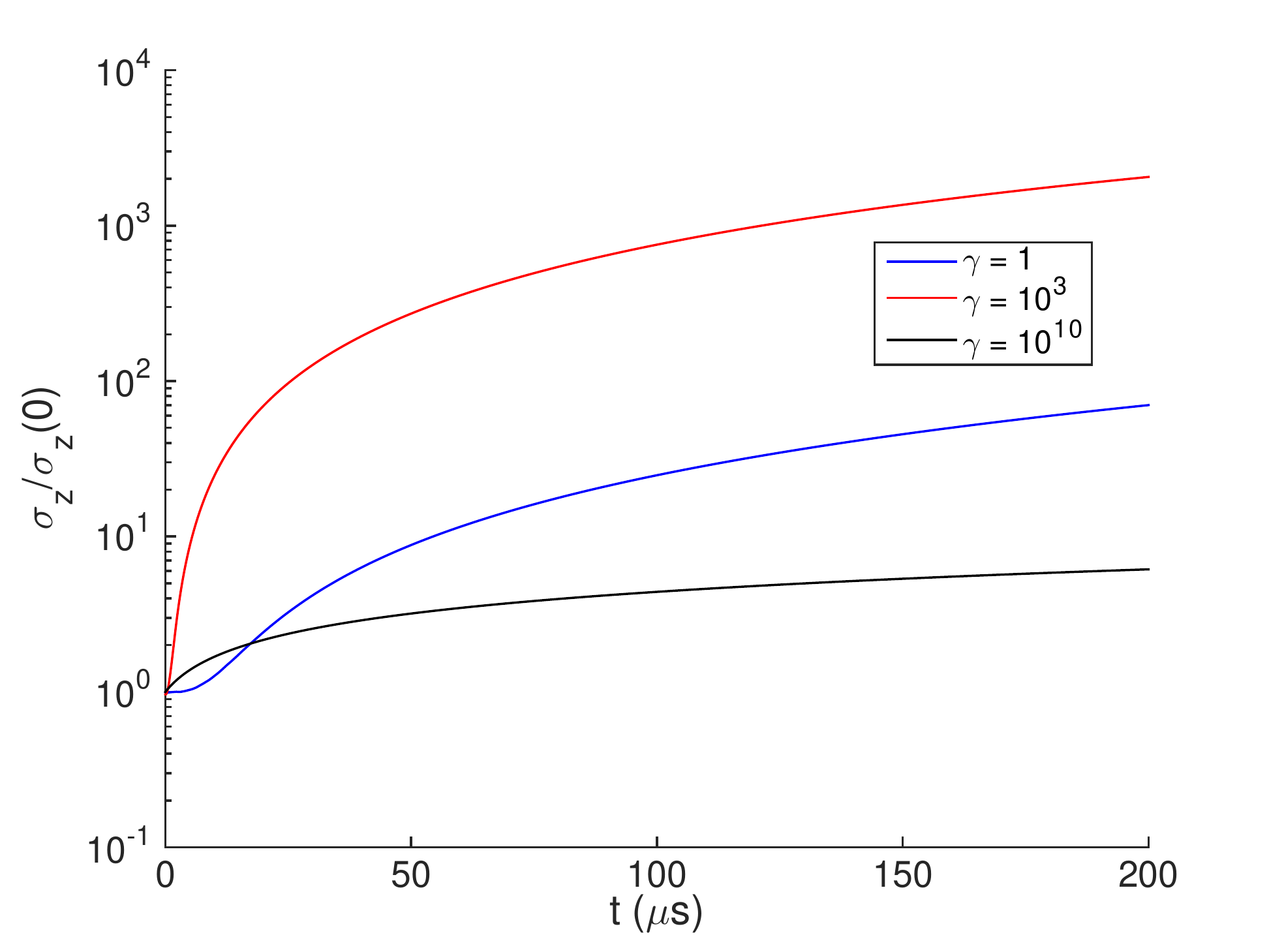}

\includegraphics[width=8cm]{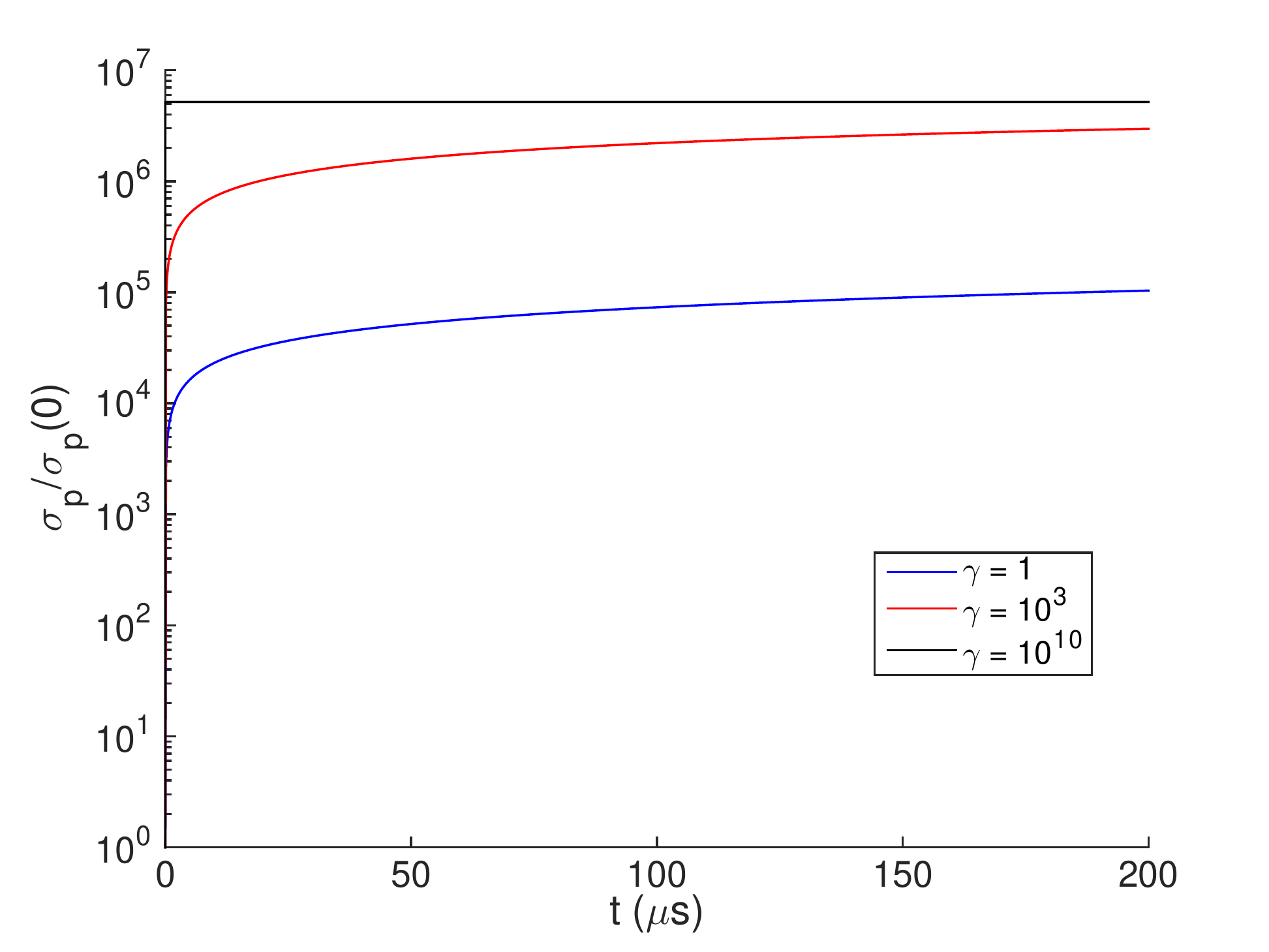}

\protect\caption{(Color online) Plots of the width of the position (up) and momentum
(down) PDFs, for three values of the parameter $\gamma$. }

\label{figsigmazandsigmap}
\end{figure}

\end{enumerate}
\begin{figure*}
\includegraphics[width=150pt]{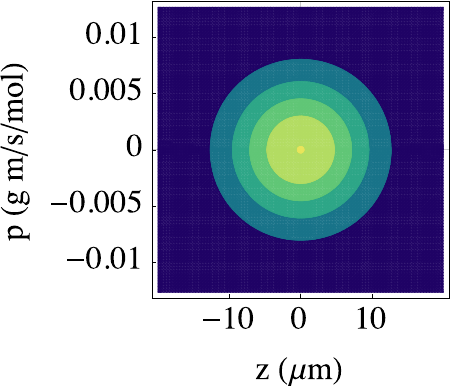} \includegraphics[width=150pt]{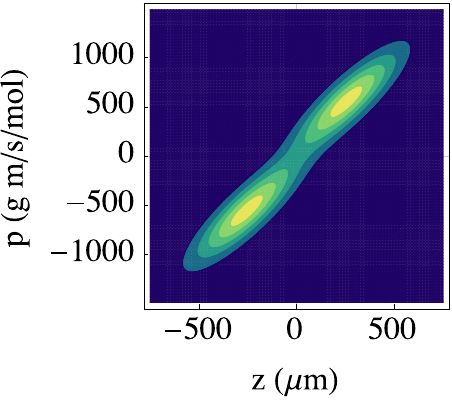}
\includegraphics[width=150pt]{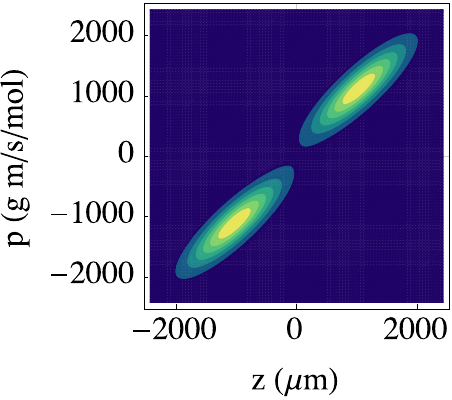}

\protect\caption{\label{fig:gamma1} (Color online) Contour plots of the trace of the
matrix Wigner function with $\gamma=1$ s$^{-1}$. The left panel
corresponds to the initial ($t=0)$ state, while the middle panel
shows the situation at $t=100\,\,\mu s$, and the right panel is for
$t=200\,\,\mu s$.}
\end{figure*}
\begin{figure*}
\includegraphics[width=150pt]{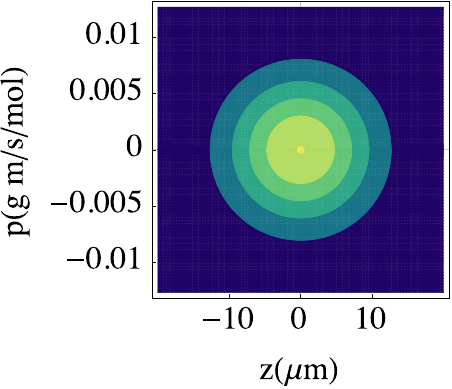} \includegraphics[width=150pt]{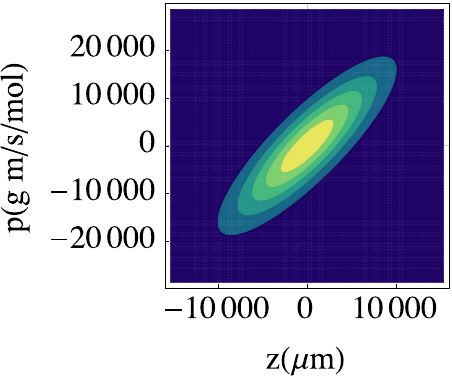}
\includegraphics[width=150pt]{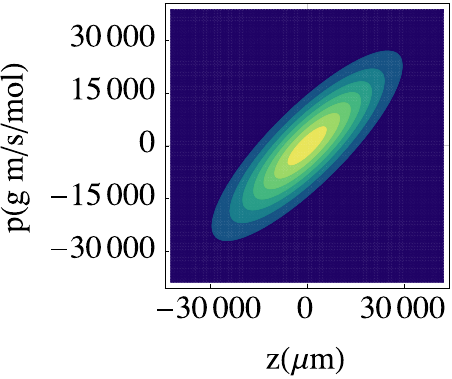} \protect\caption{\label{fig:gamma3} (Color online) Same as Fig. \ref{fig:gamma1},
for a value $\gamma=10^{3}$ s$^{-1}$. }
\end{figure*}

\begin{figure*}
\includegraphics[width=150pt]{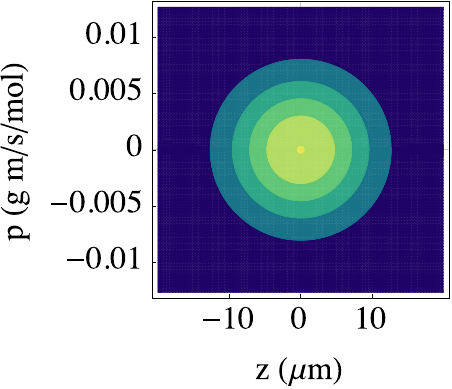} \includegraphics[width=150pt]{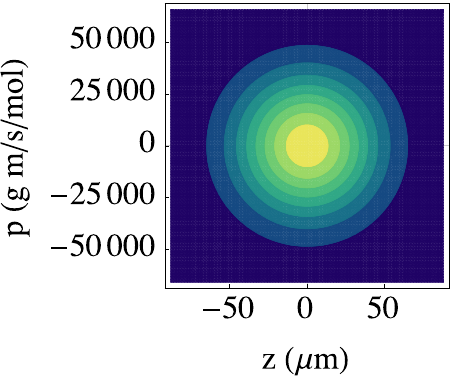}
\includegraphics[width=150pt]{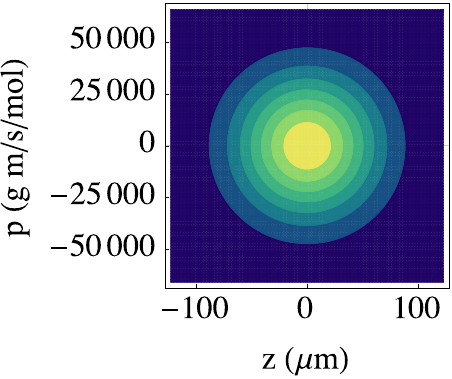} \protect\caption{\label{fig:gamma10} (Color online) Same as Fig. \ref{fig:gamma1},
using $\gamma=10^{10}$ s$^{-1}$. }
\end{figure*}

\section{\label{application}Application to the Stern-Gerlach experiment}

The study of a realistic SG experiment setup with decoherence effects
will be the core of this section. First, parameters for the setup
will be introduced. Then, the evolution of the system will be pictured
in phase space using the results from the previous section. To conclude,
the characteristic decoherence time of the system will be studied
through the damping of the off-diagonal elements of the Wigner function
as the system evolves.

\subsection{\label{parameters} Experiment parameters}

We assume an incident beam of Silver atoms ($m=1.8\times10^{-25}$
kg, $g_{s}\simeq2$) starting in the $\ket{S_{x}=\hbar/2}=\frac{1}{\sqrt{2}}\ket ++\frac{1}{\sqrt{2}}\ket -$
spin state, with an average speed $v=500$ m/s, and a beam width $\sigma=10^{-5}$
m. The SG apparatus parameters are based on the realistic ones used
in a previous work \cite{Gondran2013}. In this setup, the applied
magnetic field is $B_{0}=5$ T, and the gradient $\eta=1000$ T/m.
The longitude of the tube is $l=0.2$ m, which implies a flight time
of around $0.4$ ms for the above Silver atoms speed. This is, therefore,
the characteristic time scale for the system dynamics. The operational
temperature of the tube is around $T\simeq300$ K, i.e. the laboratory
temperature.

\subsection{\label{phase} Phase space representation}

In this section we will illustrate the behavior of the SG experiment
in a decoherent environment, by showing some plots of the Wigner function
using the parameters defined in Sect. \ref{parameters}.

\subsubsection{\label{diagonal}Diagonal elements}

In order to represent the phase space distribution of the particle,
the trace of the matrix Wigner function is plotted for $\gamma=1,\ 10^{3},\ 10^{10}$
s$^{-1}$ in Figures \ref{fig:gamma1}, \ref{fig:gamma3} and \ref{fig:gamma10},
respectively. The trace allows us to show the total quasi-probability
distribution, thus putting on the same plot both spin components.
Fig. \ref{fig:gamma1} ($\gamma=1$ s$^{-1}$) shows how the incoming
state splits into two separating components within the experiment
timescale ($<$ 1 ms): At $t=100\,\,\mu s$ both terms start to split,
and at $t=200\,\,\mu s$ they are visibly separated. We observe the
distortion of the original shape of the Wigner function, caused by
the different evolution in $z$ and $p$, that appears even when the
interaction with the environment is not included (see \cite{Utz2015SG}).
Note also that friction with the environment quickly broadens the
spatial width, from the initial value to the millimeter scale in just
$t=200\,\,\mu s$. The same behavior can be observed for the momentum
width.

As can be seen in Fig. \ref{fig:gamma3} ($\gamma=10^{3}$ s$^{-1}$),
a higher value of the friction force quickly limits the momentum of
the peak centers, and causes the growth of the distribution widths.
The influence of the SG apparatus is still visible in the shape of
the beam, but due to the speed limit and the continuous growth of
the width, the two components of the beam do not separate. An even
larger value of the friction force (Fig. \ref{fig:gamma10}) causes
the distribution peak centers to remain at the origin, and the width
of the momentum distribution quickly grows, thus completely masking
all the effects of the apparatus.

\subsubsection{\label{off-diagonal}Off-diagonal elements}

\begin{figure}
\includegraphics[width=1\linewidth]{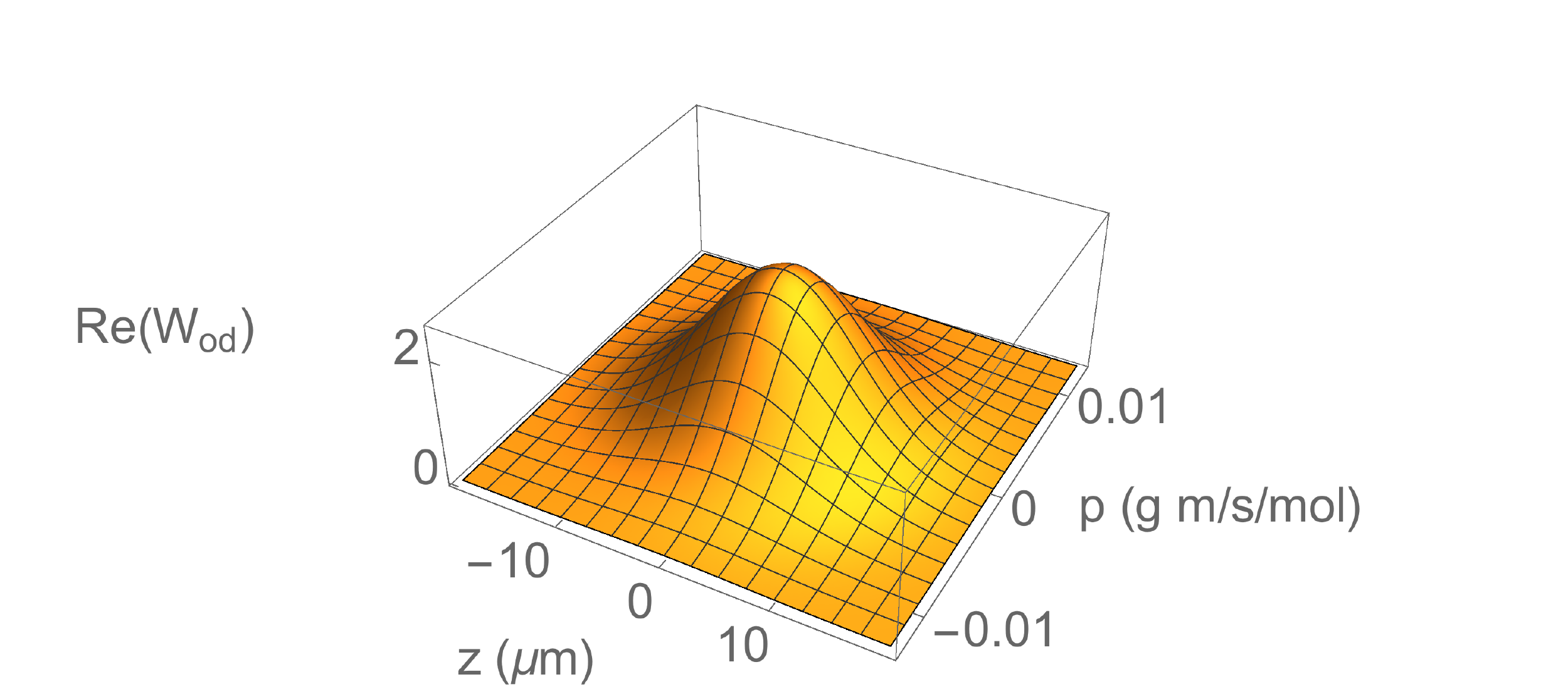}

\protect\caption{\label{fig:decogauss} (Color online) 3D-plot of the real part of
$W_{od}(z,p,t)$ at $t=0$.}
\end{figure}

To study the off-diagonal Wigner function, two 3D-plots of the real
part of $W_{od}(z,p,t)$, corresponding to $\gamma=1,\ 10^{10}$ s$^{-1}$,
are drawn in order to see how the value of the damping constant affects
the decoherence rate. At $t=0$ (Fig. \ref{fig:decogauss}) this amounts
to representing the initial Gaussian distribution Eq. (\ref{initial}).

\begin{figure}
\includegraphics[width=1\linewidth]{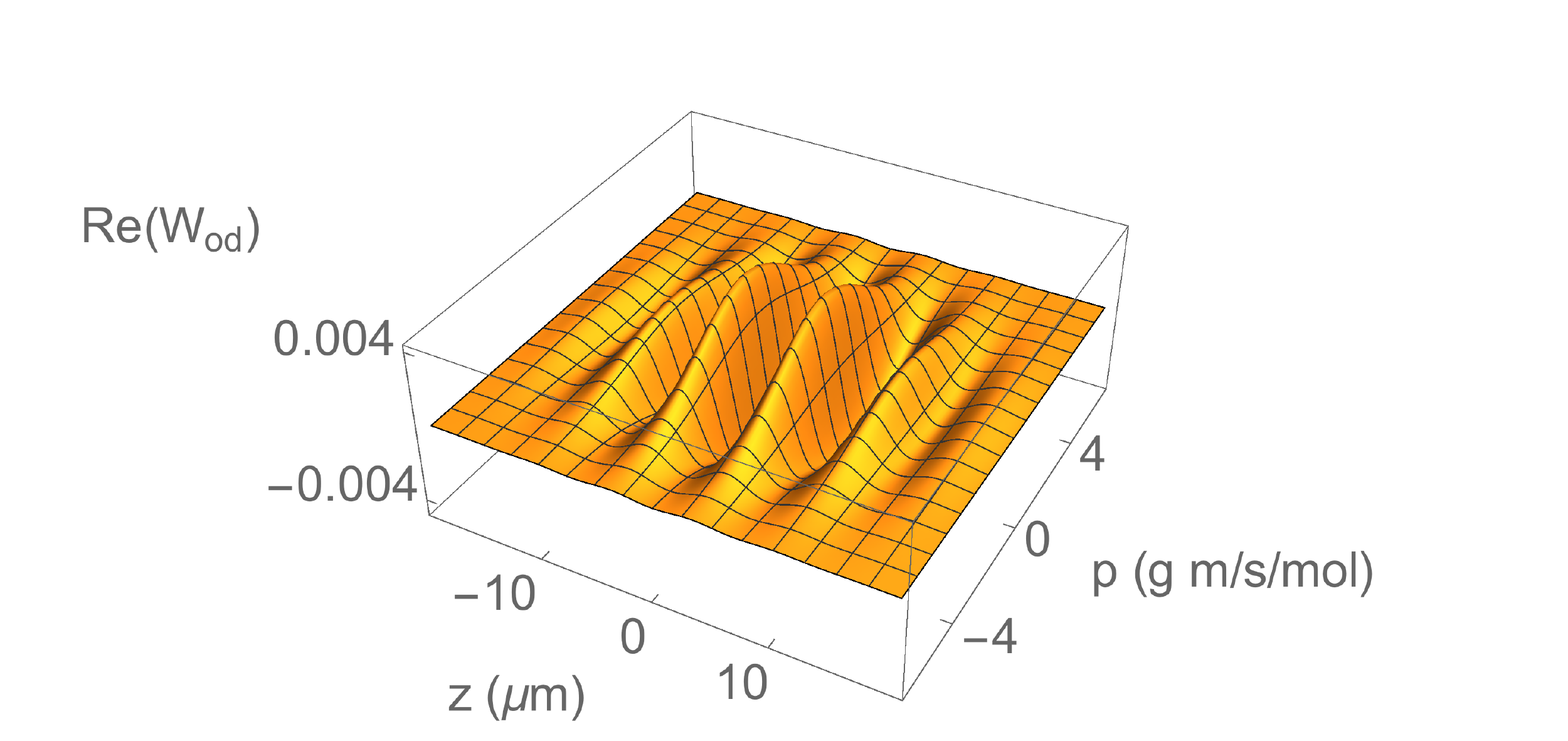} \includegraphics[width=1\linewidth]{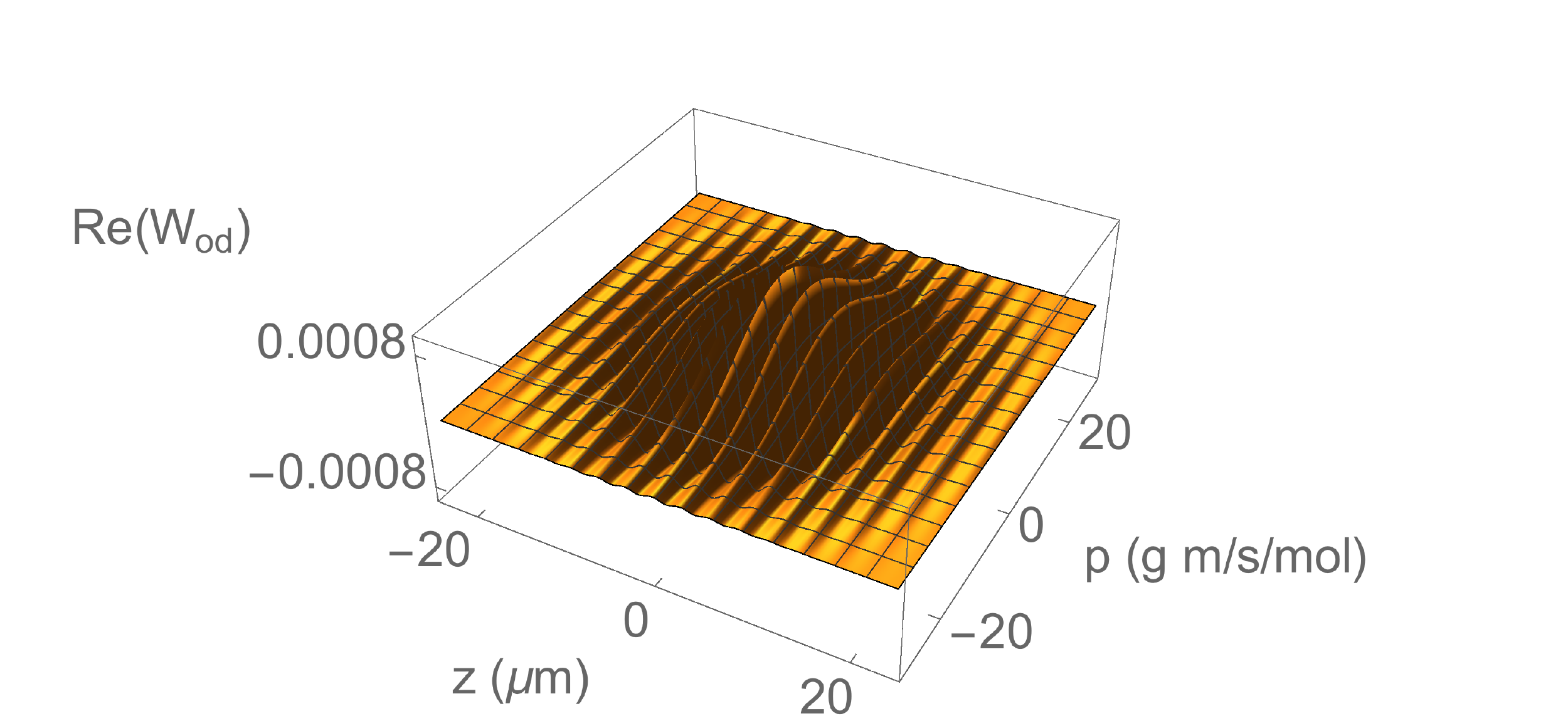}
\protect\caption{\label{fig:deco1} (Color online) 3D-plot of the real part of $W_{od}(z,p,t)$
at $t=0.005\mu s$ (up) and $t=0.1\mu s$ (down), for $\gamma=1$
s$^{-1}$.}
\end{figure}

\begin{figure}
\includegraphics[width=1\linewidth]{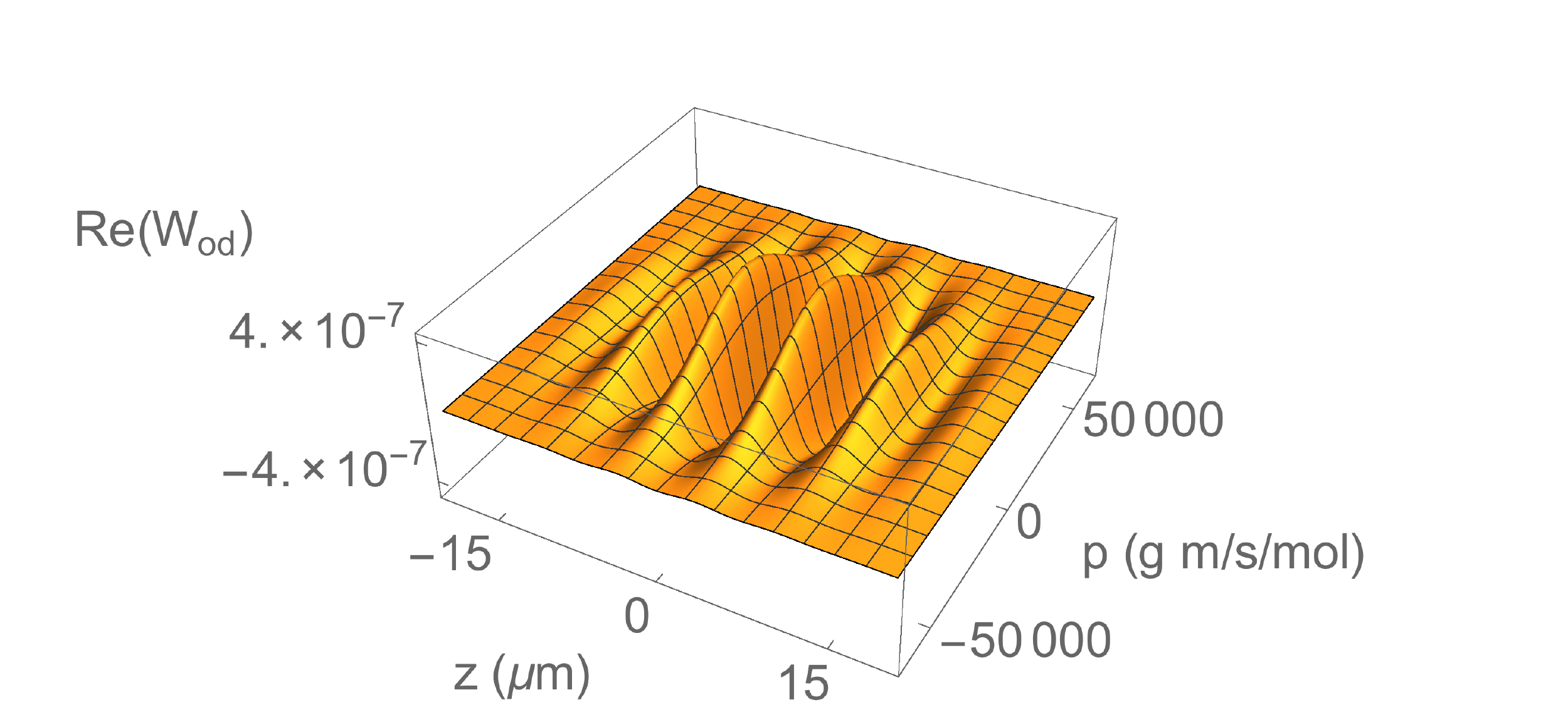} \includegraphics[width=1\linewidth]{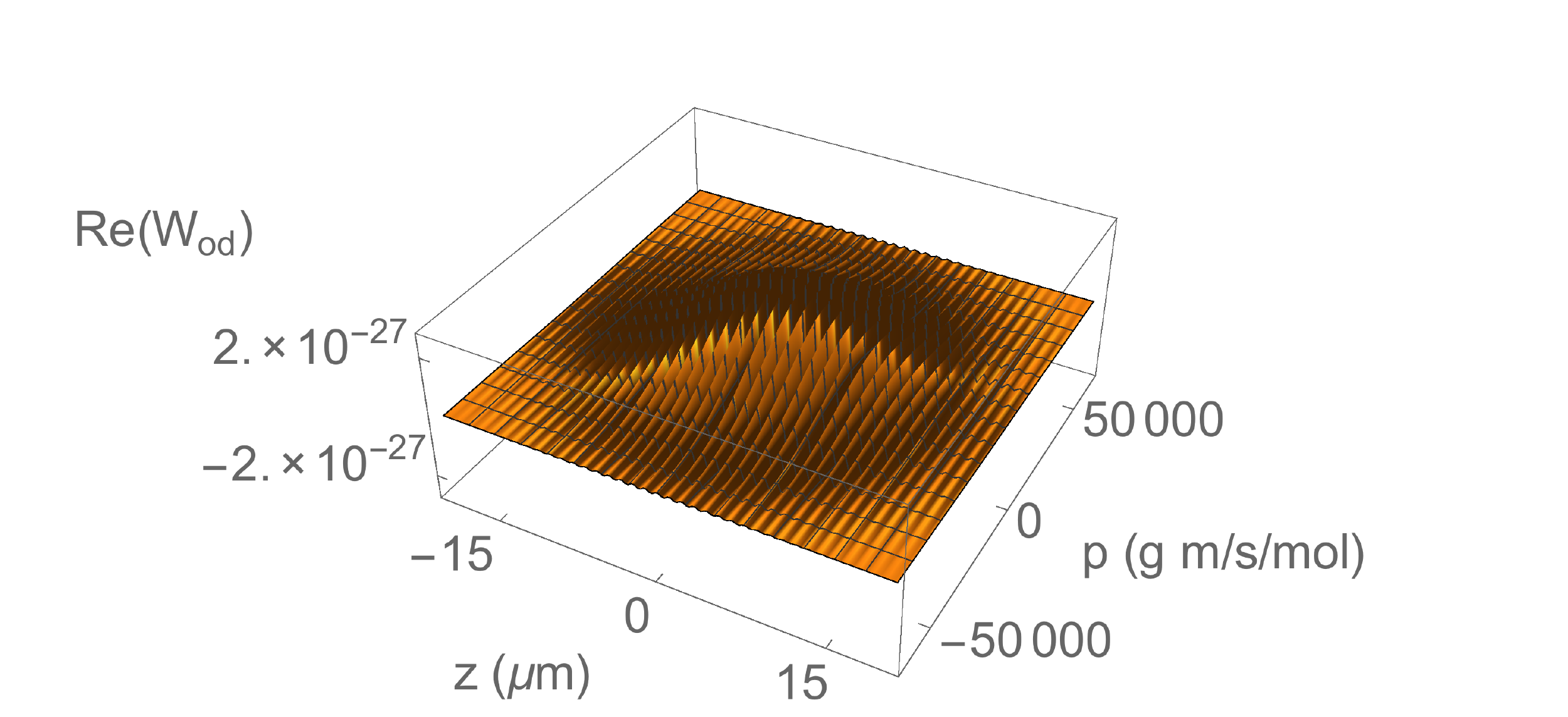}
\protect\caption{\label{fig:deco2} (Color online) Same as Fig. \ref{fig:deco1}, for
$\gamma=10^{10}$ s$^{-1}$.}
\end{figure}

As time goes on, Figs. \ref{fig:deco1} and \ref{fig:deco2} show
how the real part of $W_{od}(z,p,\tau)$ evolves, from a Gaussian
distribution, to an oscillatory function. We also observe that these
oscillations are damped due to the interaction with the environment.
By comparing both figures, we immediately see that a larger value
of $\gamma$ increases the oscillation frequency and reduces the amplitude
of the oscillations, thus leading to a faster decoherence. 

\subsection{\label{decoherence} Decoherence time}

\begin{figure}
\includegraphics[width=1\linewidth]{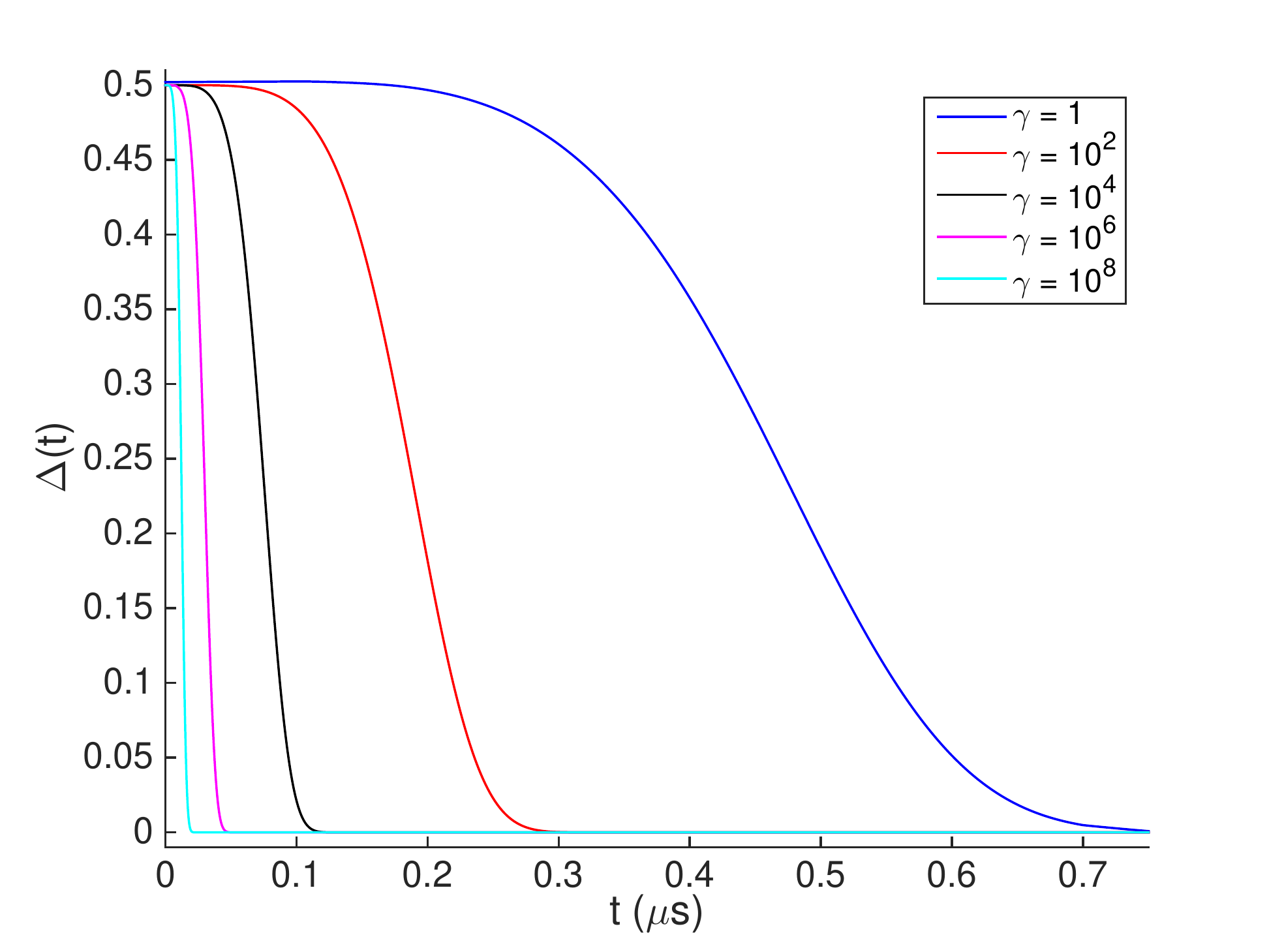} \protect\caption{\label{fig:normoff} (Color online) Plot of $\Delta(t)$, as defined
by Eq. (\ref{eta(t)}), for different values $\gamma=1,10^{2},10^{4},10^{6},10^{8}$
s$^{-1}$, as a function of time.}
\end{figure}

In order to quantify the loss of coherence in the system we choose,
as a figure of merit, the norm $\Delta(t)$ of $W_{od}(z,p,t)$ at
a given $t$, defined as 
\begin{equation}
\Delta(t)\equiv\int_{-\infty}^{\infty}\int_{-\infty}^{\infty}|W_{od}(z,p,t)|\,\,dz\,\,dp.\label{eta(t)}
\end{equation}
This quantity provides the total volume, in phase space, occupied
by the coherent term of the Wigner function. Taking the module, instead
of the function itself, avoids for cancellations due to the oscillatory
nature of this term. In Fig. \ref{fig:normoff} we have plotted $\Delta(t)$
for different values of $\gamma$. As expected, decoherence effects
manifest in a decrease of $\Delta(t)$ with time. In fact, the approach
to zero appears at earlier times as one considers larger values of
$\gamma$, clearly indicating a faster decoherence process. 

\begin{figure}
\includegraphics[width=1\linewidth]{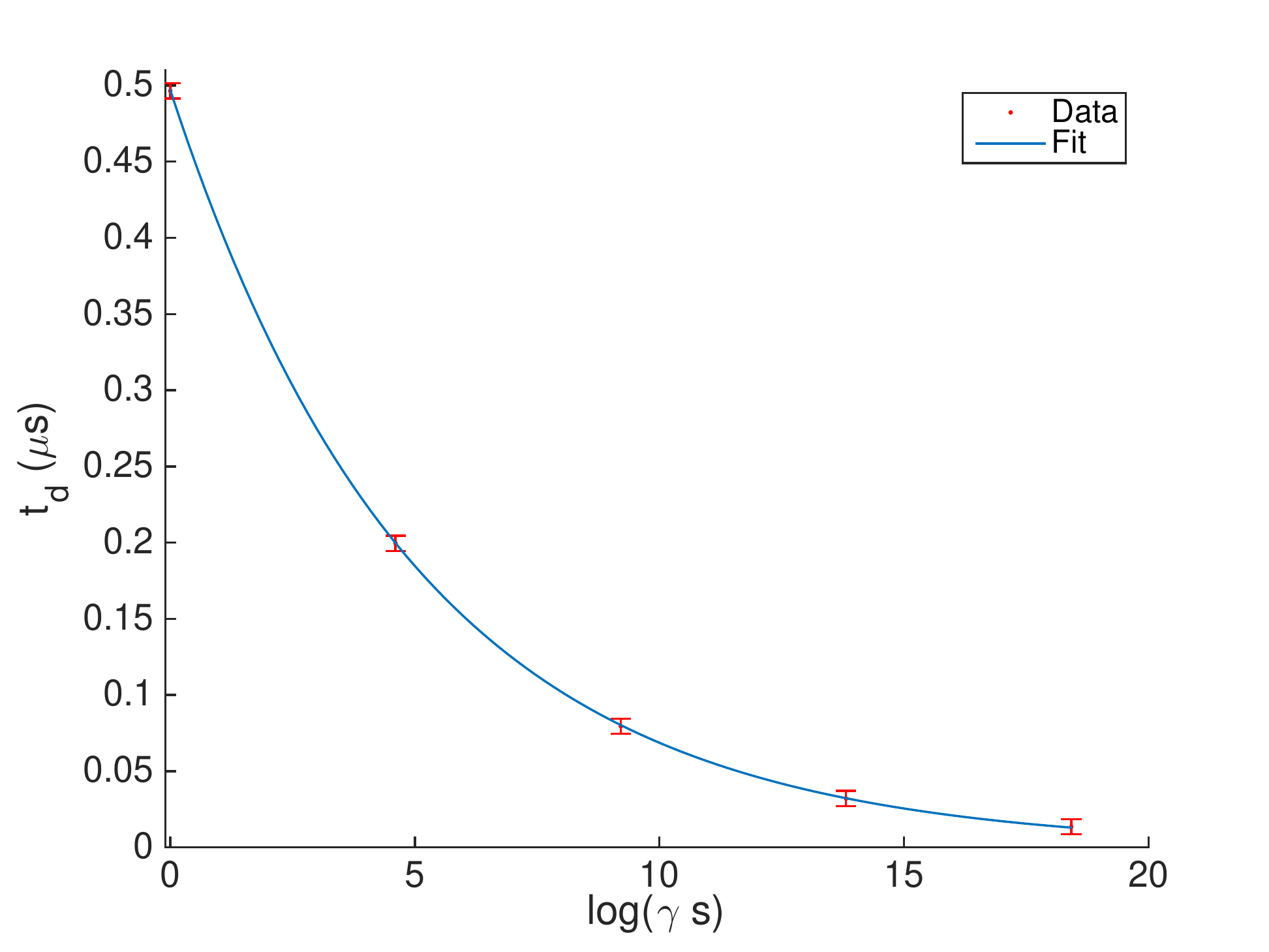} \protect\caption{\label{fig:decoherence} (Color online) Red points show the decoherence
time $t_{d}$ for the same values of $\gamma$ used in Fig. (\ref{fig:normoff}).
The bar indicates the imprecision that originates from the data. The
solid curve is our fit $t_{d}=a\ \gamma^{b}$ (see the text for explanation).}
\end{figure}

In view of this result, we can introduce a decoherence time $t_{d}$,
as the time it takes for $\Delta(t)$ to reduce its initial value
by a factor $e$. From the above data, we can obtain $t_{d}$ for
a given value of $\gamma$. These data are collected in Fig. \ref{fig:decoherence}. 

Performing a numerical fit to the curve $t_{d}(\gamma)$ we find that
the decoherence time can be approximated by the formula $t_{d}=a\ \gamma^{b}$,
with $a=0.497\pm0.002$ and $b=-0.198\pm0.001$, pointing to a behavior
as $t_{d}\propto\gamma^{-1/5}$. This result is at variance with the
one discussed in \cite{1995PhyA..220..563V}, where a decoherence
time given by $\left(\frac{3\hbar^{2}m^{2}\gamma^{2}}{4D\eta^{2}\lambda^{2}}\right)^{1/3}$
was claimed, which translates into a dependence of the form $t_{d}\propto\gamma^{1/3}$,
i.e. the larger the decoherence parameter $\gamma$, the later the
system experiences decoherence. In our opinion, such result is unrealistic,
as one expects the opposite behavior: Decoherence should take place
faster as $\gamma$ is increased, in accordance to our results.

\subsection{\label{validity} Discussion about the validity of the model}

In the SG experiment, Silver atoms travel through the high vacuum
beam pipe. We assume that they collide with the residual air molecules,
resulting into Brownian motion and decoherence. In the Caldeira-Leggett
master equation, these effects are included via the damping rate $\gamma$.
The Langevin equation relates $\gamma$ to the viscosity $\mu$ as
follows \cite{Zurek2003} 
\begin{equation}
\gamma=\frac{\mu}{2m},
\end{equation}
with $\mu$ the viscosity of the medium for Silver atoms. This quantity
can be explicitly calculated through the following expression \cite{Marquardt1999}
\begin{equation}
\mu=6\pi R\mu'=2.994\ \pi RnM_{M}v_{M}\lambda_{M},
\end{equation}
where $R=144$ pm is the radius of the Silver atom, $n$ is the particle
number density of the air; $M_{M}$, $v_{M}$ and $\lambda_{M}$ are
the molecular mass, mean speed and mean free path of air molecules,
respectively. The Stokes' law term $6\pi R$ accounts for the contribution
of Silver atoms mass to the viscosity \cite{Batchelor1968}, and the
term $\mu'=0.499\ nM_{M}v_{M}\lambda_{M}$ accounts for the air contribution.
Assuming typical values for the pressure of the beam pipe $P\simeq10^{-6}$
mbar \cite{PHYWE}, one finds \cite{Marquardt1999} $n\simeq10^{10}$
cm$^{-3}$, $\lambda_{M}\simeq10^{4}$ cm, and $v_{M}\simeq500$ m/s.
To obtain a rough estimate, we assume that the molecules in the low
pressure conditions in the tube are mostly composed by $N_{2}$, with
a mass $M_{M}\simeq28$ uma $=4.65\ 10^{-26}$ kg.

Using the above data, one obtains the estimate $\mu\sim10^{-14}$
Kg/s, implying that the order of magnitude of $\gamma$ is $\sim10^{10}$
s$^{-1}$. Such value, however, posses a problem: As we have seen
(Fig. \ref{fig:gamma3}), for damping rates larger than $10^{3}$
s$^{-1}$, the two components of the initial state do not separate
due to the high friction with the environment. This is obviously not
what happens in the real experiment, leading us to serious doubts
of the application of the Caldeira-Leggett model to the SG experiment.

A possible explanation of this disagreement might be that the obtained
value of $\gamma$ is wrong, because the calculation of the viscosity
discussed above is not correct in high vacuum. As \cite{Marquardt1999}
indicates, when $\lambda_{M}$ is larger than the size of the container,
the gas is in a molecular state, and cannot be characterized by a
viscosity anymore. Furthermore, as explained in \cite{PHYWE}, due
to the high vacuum in the beam pipe, the mean free path of Silver
atoms is a multiple of the beam pipe length. Under these conditions,
one should question the validity of the master equation Eq. (\ref{edo-1}),
which was used to derive the corresponding differential equations
that describe the Wigner function dynamics.

\section{\label{conclusions}Conclusions}

In this paper, we have studied the SG experiment with environmental
induced decoherence described by the Caldeira-Leggett model. Our description
is done on the phase space, making use of Wigner functions, with the
additional spin degree of freedom. Our goal was to describe the kinematics
of the atoms traversing the magnetic field gradient, and interacting
with a thermal bath of particles, leading to Brownian motion and decoherence.
We solved the differential equations for the Wigner function corresponding
to the model, starting from an initial separable state, with a Gaussian
shape in space, and an arbitrary spin state. The diagonal terms on
the $S_{z}$ basis have a simple interpretation, in terms of the two
separating states of the apparatus. By calculating the marginals over
momentum or position for each of the diagonal terms of the Wigner
function, we obtain the probability distribution for the conjugate
variable. Each of the obtained diagonal distributions conserve the
initial Gaussian shape, both in position and in space, in spite of
the interaction with the environment, and allow for a clear description
in terms of the center and width of the corresponding Gaussian, which
bear a close analogy with the classical Brownian motion of a particle.
In particular, at large times the particle reaches a limit velocity,
a feature which is particularly important for the description of the
SG experiment inside a gas tube. We also showed that our results agree
with the non decoherence expressions in the appropriate limit. 

By adopting realistic parameters for the SG experiment, we plotted
the diagonal and off-diagonal elements of the Wigner function. We
observe that, for low values of the decoherence parameter $\gamma$,
the initial state separates into two components, corresponding to
the up and down spin, as expected. The off-diagonal terms provide
us information about decoherence, as a consequence of the interaction
with the environment, that manifests in the damping of these terms
as time evolves. In order to quantify the effect of decoherence, we
evaluate the norm of the off-diagonal element over the whole phase
space. The characteristic decoherence time scale $t_{d}$ is defined
as the instant when the initial norm is reduced by a factor $e$.
By obtaining $t_{d}$ for different values of $\gamma$, we find a
relation which is well described by a power law $t_{d}\propto\gamma^{-1/5}$.
Our result differs from previous results, where a dependence $t_{d}\propto\gamma^{1/3}$
was claimed instead. We argue that such dependence is counter intuitive,
as it would imply that a larger value of $\gamma$ makes decoherence
effects to appear later. In contrast, we obtain that a stronger value
of $\gamma$ implies a shorter decoherence time, which seems more
reasonable.

Finally, we discuss a caveat of the model, that shows up after estimating
the order of magnitude of $\gamma$ in a realistic scenario. In fact,
we obtain $\gamma\sim10^{10}$, a value that would lead to a large
friction, and would ruin the observed phenomenology of the experiment.
This estimate has itself a loophole, as the calculated mean free path
of the incoming particles on the tube turns out to be of the order
of the tube length, thus implying that collisions with the air molecules
are rare events. Altogether, these reasonings rise some doubts on
the applicability of the Caldeira-Leggett model for this problem.
In our opinion, this question deserves further research, given the
importance of the experiment as one of the cornerstones in quantum
physics.
\begin{acknowledgments}
This work has been supported by the Spanish Ministerio de Educación
e Innovación, MICIN-FEDER projects FPA2011-23897 and FPA2014-54459-P,
and ``Generalitat Valenciana'' grant GVPROMETEOII2014-087. We gratefully
acknowledge useful conversations with J.A. Manzanares.
\end{acknowledgments}

\section{Auxiliary formulae for the diagonal and off-diagonal Wigner function}

In this Appendix, we give the explicit expressions for auxiliary functions
that define both the diagonal and off-diagonal terms of the matrix
WF. The functions $F(z,p,\tau)$ and $G(\tau)$ that enter in the
diagonal part Eq. (\ref{wdiag}) are defined as
\begin{widetext}
\begin{widetext}
\begin{eqnarray}
F(z,p,\tau) & = & 2\gamma D\sigma^{2}\left[\left(\gamma^{2}m^{2}\left(1-e^{-2\tau}\right)z'^{2}-2\gamma mp'\left(1-e^{-\tau}\right)^{2}z'+p'^{2}\left(-\left(2-e^{-\tau}\right)^{2}+2\tau+1\right)\right)\right.\nonumber \\
 & + & \left.\gamma^{4}m^{2}p'^{2}\sigma^{4}+\gamma^{2}\hbar^{2}\left(p'\left(1-e^{-\tau}\right)-\gamma me^{-\tau}z'\right)^{2}\right],
\end{eqnarray}
\begin{eqnarray}
G(\tau) & = & 8D^{2}\sigma^{2}\left(1-e^{-\tau}\right)\left(\left(e^{-\tau}+1\right)\tau-2\left(1-e^{-\tau}\right)\right)+2\gamma D\left[\gamma^{2}m^{2}\sigma^{4}\left(1-e^{-2\tau}\right)\right.\nonumber \\
 & + & \left.\left(e^{-2\tau}(2\tau+3)-4e^{-\tau}+1\right)\hbar^{2}\right]+\gamma^{4}m^{2}\sigma^{2}e^{-2\tau}\hbar^{2},
\end{eqnarray}

\end{widetext}

\end{widetext}

where $z'=z\mp z_{c}$, $p'=p\mp p_{c}$. The magnitudes $z_{c}$
and $p_{c}$ are given in Eqs. (\ref{zc}) and (\ref{pc}), respectively. 

For the off diagonal term defined in Eq. (\ref{woff}) we introduced
the following definitions:
\begin{widetext}
\begin{equation}
C_{1}(\tau)=\frac{\eta^{2}\lambda^{2}\left(\sigma^{2}\left(-2D\left(2\tau^{3}-6\tau^{2}+6\tau+3\right)+6De^{-2\tau}+24De^{-\tau}\tau-3\gamma^{3}m^{2}\sigma^{2}\tau^{2}\right)-3\gamma\left(1-\tau-e^{-\tau}\right)^{2}\hbar^{2}\right)}{3\gamma^{5}m^{2}\sigma^{2}\hbar^{2}},
\end{equation}
\begin{equation}
C_{2}(\tau)=\frac{\eta\lambda\left(\gamma\hbar^{2}\left(e^{-\tau}-e^{-\tau}\tau-e^{-2\tau}\right)-2D\sigma^{2}\left(1-2e^{-\tau}\tau-e^{-2\tau}\right)\right)}{\gamma^{3}\hbar^{2}m\sigma^{2}},
\end{equation}
\begin{equation}
C_{3}(\tau)=\frac{\sigma^{2}\left(2De^{-2\tau}-8De^{-\tau}-D(4\tau-6)-\gamma^{3}m^{2}\sigma^{2}\right)-\gamma\hbar^{2}\left(1-e^{-\tau}\right)^{2}}{4\gamma^{3}m^{2}\sigma^{2}},
\end{equation}
\begin{equation}
C_{4}(\tau)=\frac{\eta\lambda\left(\gamma\left(1-e^{-\tau}\right)\left(1-\tau-e^{-\tau}\right)\hbar^{2}-\sigma^{2}\left(2D\left(\tau+e^{-\tau}-1\right)^{2}+\gamma^{3}m^{2}\sigma^{2}\tau\right)\right)}{\gamma^{4}m^{2}\sigma^{2}\hbar},
\end{equation}
\begin{equation}
C_{5}(\tau)=\frac{\left(e^{-\tau}-1\right)\left(\gamma e^{-\tau}\hbar^{2}+2D\sigma^{2}\left(1-e^{-\tau}\right)\right)}{2\gamma^{2}m\sigma^{2}\hbar},
\end{equation}
\begin{equation}
C_{6}(\tau)=\frac{\gamma e^{-2\tau}\hbar^{2}+2D\sigma^{2}\left(1-e^{-2\tau}\right)}{4\gamma\sigma^{2}\hbar^{2}}.
\end{equation}
 
\end{widetext}


%
\end{document}